\def\be{\begin{equation}}
\def\ee{\end{equation}}
\def\ber{\begin{eqnarray}}
\def\eer{\end{eqnarray}}
\def\bers{\begin{eqnarray*}}
\def\eers{\end{eqnarray*}}

\documentclass[aps,prb,twocolumn,showpacs,groupedaddressamsmath,amssymb]{revtex4-2}
\usepackage{dcolumn} % Align table columns on decimal point
\usepackage{amsmath}
\usepackage{multirow}
\usepackage{setspace}
\usepackage{blindtext}
\usepackage{graphicx} % Include figure files
\usepackage{subfigure}
\usepackage{comment}
\usepackage{color}
\usepackage{makecell}
\usepackage{ifthen}
\newboolean{includefigs}
\setboolean{includefigs}{true} % Permits rapid compiling by removing figures/text
\newboolean{includetext}
\setboolean{includetext}{true} % false = removed, true = included
\newcommand{\condcomment}[2]{\ifthenelse{#1}{#2}{}}

\begin{document}

%%%%%%%%%%%%%%%%%%%%%%%%%%%%%%%%%%%%%%%%%%%%%%%%%
% TITLE
%%%%%%%%%%%%%%%%%%%%%%%%%%%%%%%%%%%%%%%%%%%%%%%%
\title{Thermoelectric properties of Topological Weyl Semimetal Cu$_2$ZnGeTe$_4$}
\author{Bhawna Sahni$^1$$^{\dagger}$, Himanshu Sharma$^5$$^{\dagger}$, Riddhimoy Pathak$^2$, P C Sreeparvathy$^3$, Tanusri Saha-Dasgupta$^4$, Kanishka Biswas$^2$ and Aftab Alam$^5$ }
\email{aftab@iitb.ac.in}
\thanks{$^{\dagger}$These authors contributed equally to this work}
\affiliation{$^1$School of Engineering, University of Warwick, Coventry, CV4 7AL, United Kingdom}
\affiliation{$^2$New Chemistry Unit, Jawaharlal Nehru Centre for Advanced Scientific Research
(JNCASR), Jakkur P.O., Bangalore 560064, India}
\affiliation{$^3$Department of Physics, National Sun Yat-sen University, Kaohsiung 80424, Taiwan}
\affiliation{$^4$Department of Condensed Matter and Materials Physics, S. N. Bose National Centre for Basic Sciences, JD Block, Sector III, Salt Lake, Kolkata, West Bengal 700106, India}
\affiliation{$^5$Department of Physics, Indian Institute of Technology Bombay, Mumbai 400076, Maharashtra, India}

%%%%%%%%%%%%%%%%%%%%%%%%%%%%%%%%%%%%%%%%%%%%%%%%%
% ABSTRACT
%%%%%%%%%%%%%%%%%%%%%%%%%%%%%%%%%%%%%%%%%%%%%%%%
\begin{abstract}
The exploration of topological quantum materials for thermoelectric (TE) applications offers an opportunity to combine nontrivial electronic topology with efficient energy conversion. Topological semimetals (TSMs), including Dirac, Weyl, and nodal-line systems, can exhibit favourable transport properties arising from the coexistence of dispersive and relatively flat bands near the Fermi level ($E_F$). However, their gapless electronic structure can also suppress thermopower and enhance electronic thermal conductivity, limiting the overall TE performance. Here, we combine first-principles calculations and transport measurements to investigate Cu$_2$ZnGeTe$_4$ as a lattice-tunable platform connecting thermoelectric and topological electronic phases. At the experimentally measured lattice parameters, Cu$_2$ZnGeTe$_4$ is a narrow-gap semiconductor with a calculated band gap of approximately 0.067 eV and a maximum calculated $ZT$ of approximately 1. Experimentally, the compound exhibits $p$-type semiconducting behavior and low lattice thermal conductivity, yielding a $ZT$ of approximately 0.14 at 623 K and reproducing the calculated temperature-dependent transport trends. Upon lattice expansion, first-principles calculations predict band inversion and a transition to a Weyl-semimetallic phase with Weyl nodes of opposite chirality and topological surface states. Although this phase exhibits enhanced electrical conductivity and a high power factor, its maximum $ZT$ of approximately 0.36 remains lower than that of the semiconducting phase because of concomitant increases in electronic and lattice thermal conductivities. Isovalent Sn substitution at the Ge site reproduces the essential electronic features of the expanded phase, providing a possible chemical route toward the predicted topologically nontrivial state. These results demonstrate that topological semimetals can retain competitive thermoelectric performance while highlighting the competing electronic and phononic transport effects of topological band engineering. Cu$_2$ZnGeTe$_4$ thus provides a tunable platform for exploring the interplay between nontrivial topology and thermoelectric transport.
\end{abstract}

\date{\today}

\maketitle

\section{Introduction}

Thermoelectric (TE) materials have long attracted interest as a means of converting waste heat into useful electrical energy.\cite{key-1, key-9} Their performance is characterised by the dimensionless figure of merit
$ZT = S^2\sigma/(\kappa_e+\kappa_L)$, where $S$, $\sigma$, $\kappa_e$, and $\kappa_L$ denote the Seebeck coefficient, electrical conductivity, electronic thermal conductivity, and lattice thermal conductivity, respectively. While hierarchical phonon scattering\cite{Heir_phn,phn_eng1} and phonon engineering\cite{phn_eng} can effectively suppress $\kappa_L$, further improvement of $ZT$ requires simultaneous optimisation of the electronic transport coefficients, which are often strongly coupled.\cite{Weidmann} Band engineering provides an effective route to partially decouple these coefficients and enhance the power factor ($S^2\sigma$).\cite{BE1,BE2}

The coexistence of thermoelectricity and nontrivial electronic topology has therefore emerged as an attractive direction for materials design. Several established thermoelectric materials, including Bi$_2$Te$_3$, Sb$_2$Te$_3$, Bi$_2$Se$_3$, and SnTe, also exhibit topological-insulator phases.\cite{TE-TI1}$^{-}$\cite{TE-TI5} Their favourable properties are associated with heavy constituent elements, narrow band gaps, and strong spin--orbit coupling, which can generate inverted bands and complex electronic structures beneficial for thermoelectric transport.\cite{Bite} Heavy elements can also contribute to reduced lattice thermal conductivity. In topological insulators, however, the transport advantages of topological surface states require the Fermi level ($E_F$) to lie within the bulk gap, whereas efficient thermoelectric transport generally relies on bulk carriers. This distinction makes bulk topological states particularly attractive for thermoelectric applications, where nontrivial topology and useful bulk transport can potentially coexist.

\begin{figure}[t]
\centering
\includegraphics[scale=0.6]{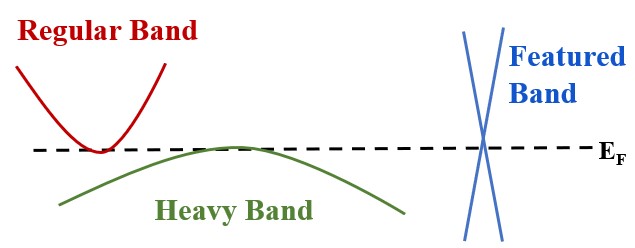}
\caption{Schematics of different band features whose coexistence are useful for promising thermoelectric (TE) performance.}
\label{conv_TI}
\end{figure}

Topological semimetals (TSMs), including Dirac and Weyl semimetals,\cite{Dirac_Weyl} offer an alternative platform in which topological bulk states coexist naturally with conducting surface states. Their gapless electronic structures can, however, lead to small thermopower through bipolar transport, limiting their thermoelectric performance. Nevertheless, recent studies have demonstrated promising TE properties in selected semimetals.\cite{Co_exist,sm1} A favourable scenario arises when relatively flat and highly dispersive bands coexist near $E_F$ (see Fig.~\ref{conv_TI}). The dispersive bands provide low effective masses and high carrier mobility, whereas the flatter bands enhance the density of states and Seebeck coefficient. Such a combination can yield a large power factor while mitigating the limitations associated with conventional semimetallic electronic structures. This band-feature coexistence makes Dirac and Weyl semimetals promising candidates for thermoelectric applications. At the same time, nontrivial topology does not intrinsically guarantee a higher $ZT$ than that of a conventional semiconductor: enhanced band velocities can increase electrical conductivity but may also increase $\kappa_e$, while the gapless spectrum can reduce $S$ through electron--hole compensation. Thus, an important materials-design objective is to access a topologically nontrivial phase while retaining a competitive power factor and low thermal conductivity.

Chalcogenides are particularly attractive for realizing this combination of electronic and thermal properties because of their broad applications in solar energy,\cite{chalco1} batteries,\cite{chalco2} thermoelectric devices,\cite{chalco3} and topological insulators.\cite{TE-TI1} Quaternary chalcogenides have, in particular, emerged as promising materials for photovoltaics and electrocatalysis,\cite{chalco56} as well as for realizing topological insulating phases.\cite{chalco7} Despite extensive studies of their structural, electronic, and functional properties, the interplay between thermoelectric transport and nontrivial topology in these materials remains comparatively unexplored. Moreover, many quaternary chalcogenides exhibit relatively large band gaps, such as 1.41~eV for Cu$_2$ZnSnSe$_4$ and 0.98~eV for Cu$_2$CdSnSe$_4$,\cite{wide-band} which can limit their electronic transport compared with conventional thermoelectric materials. Heavy and relatively flat bands can further reduce carrier mobility and electrical conductivity.

In this work, we investigate Cu$_2$ZnGeTe$_4$ as a lattice-tunable platform connecting thermoelectric transport with nontrivial electronic topology. At the experimentally measured lattice parameters, Cu$_2$ZnGeTe$_4$ is a narrow-gap semiconductor with a calculated band gap of 0.067~eV and a maximum calculated $ZT$ of approximately 1. Our experimental measurements establish the synthesis, $p$-type transport, narrow-gap semiconducting character, and low lattice thermal conductivity of the parent compound, providing a qualitative comparison with the calculated temperature-dependent transport properties. These measurements validate the thermoelectric relevance of the experimentally accessible semiconducting phase but do not directly probe the predicted topological phase. Upon lattice expansion, first-principles calculations reveal band inversion and a transition to a Weyl-semimetallic phase hosting Weyl nodes of opposite chirality and topological surface states. Although the Weyl phase exhibits enhanced electrical conductivity and a high power factor, its maximum $ZT$ is lower than that of the semiconducting phase because of concomitant increases in electronic and lattice thermal conductivities. We further examine isovalent Sn substitution at the Ge site, i.e. Cu$_2$ZnGe$_{1-x}$Sn$_x$Te$_4$, as a chemical route to induce the required lattice expansion and access the predicted topologically nontrivial phase. A comparison between the calculated and measured transport properties of the experimentally realized semiconducting phase is presented in Sec.~III~B.

\section{Methods}
%\subsection{Computational Details} 
First-principles calculations were performed using Vienna \textit{Ab-initio} simulation package (VASP)\cite{VASP1,VASP2,VASP3} with a projector augmented wave\cite{PAWbasis} basis and the generalized gradient approximation (GGA)\cite{GGA} exchange-correlation functional of Perdew$-$Burke$-$Ernzerhof (PBE).\cite{PBE01,PBE02} 
Further Computational details are provided in the Supplementary Material (SM).\cite{suppl}

 Copper (Ag, Strategic Metals Investment Ltd., 99.9\%), Zinc (Zn, Alfa Aesar, 99.9 \%), tellurium (Te, Strategic Metals Investment Ltd, 99.99 \%) and germanium (Ge, Strategic Metals Investment Ltd, 99.999 \%) were used for synthesis of Cu$_2$ZnGeTe$_4$. High-quality crystalline ingots ($\sim$4 g) of Cu$_2$ZnGeTe$_4$ were synthesized by combining stoichiometric amounts of high-purity Cu, Zn, Ge, and Te in a quartz ampoule (10 mm diameter). Further details about synthesis, powder XRD, thermal conductivity, electrical trasnport, Hall and band gap measurements are provided in the SM.\cite{suppl}
 
 %A plane wave cutoff of 500 eV is used for the calculations. The Brillouin zone sampling was done by using a $\Gamma$-centered $k$-mesh. For simulations, a k-mesh of 8$\times$8$\times$4 (for ionic relaxations) and 12$\times$12$\times$6 (for self-consistent calculations) were used. Modified Becke Johnson (mBJ) \cite{mBJ} exchange-correlation functional including spin-orbit coupling effect was performed for the accurate estimation of band gaps. The phonon dispersion and second-order inter-atomic force constants (IFCs) are calculated using phonopy code.\cite{phonopy} The third-order IFCs are calculated by taking the conventional unit cell. Using these IFCs, the lattice thermal conductivity is calculated using Phono3py code.\cite{Phono3py} AMSET \cite{AMSET} was used to calculate the thermoelectric (TE) properties of these materials. Topological properties of Cu$_{2}$ZnGeTe$_{4}$ were calculated by combing VASP package and Wannier90\cite{wannier90}, Wannier-Tool\cite{wt} packages. Topological properties such as surface state and Berry curvature arc are reported.

%\textcolor{red}{Experimental synthesis: ??}
\section{Results and discussions}
\subsection{Theoretical Results}
\subsubsection{Crystal structure}

\begin{table}[b]
    \centering
    \small % Slightly reduces font size to ensure it fits perfectly in one column
    \setlength{\tabcolsep}{0.005pt}
    \begin{tabular}{|c|c|c|c|c|}
        \hline 
        Lattice & Theory & Theory & Measured & Measured \\
         parameters &(This work) & (Others\cite{theor}) &  (This work) & (Others\cite{exp_a,exp_b}) \\
        \hline 
        \hline 
        a = b (\AA) & 6.10 & 6.09 & 5.95& 5.99; 5.95 \\ \hline 
        c (\AA)     & 12.09 & 12.22 & 11.89 & 11.92; 11.85 \\ \hline 
     %   V (\AA$^3$) & 449.87 & 453.22 & 420.94 & 427.69; 419.52 \\ \hline 
                
    \end{tabular}
	\caption{Theoretically optimized and experimental measured lattice parameters of Cu$_{2}$ZnGeTe$_{4}$. Data from other published articles are also given for comparison. } 
	\label{lattice}
\end{table}

\begin{figure*}[!t]
	\centering
	\includegraphics[scale=0.19]{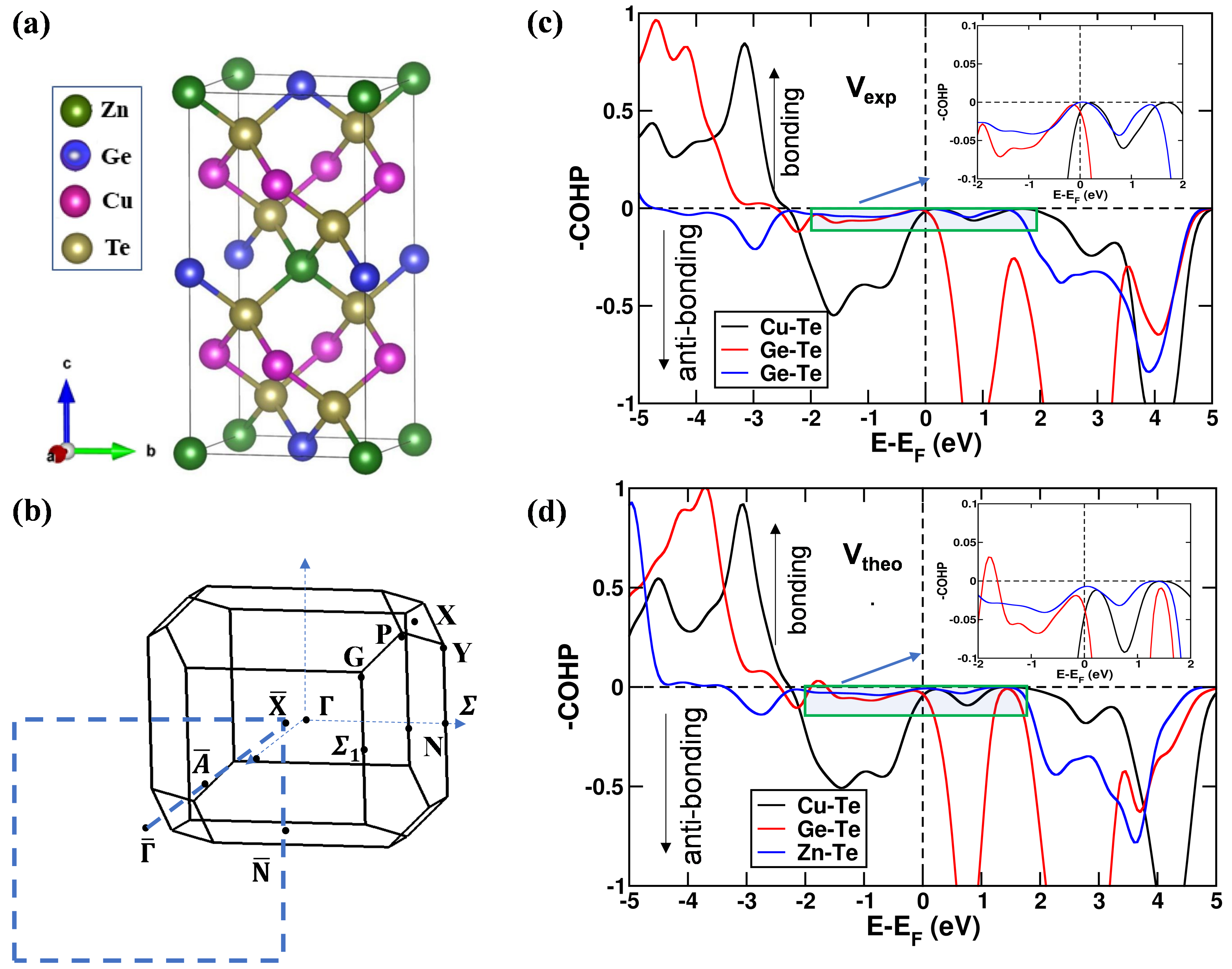}
	\caption{For Cu$_2$ZnGeTe$_4$, (a) crystal structure with space group I$\bar{4}$2m ($\#$121) (b) Bulk and (001) surface Brillouin zones.  (c,d) Crystal orbital Hamilton population (COHP) for the structures at experimental volume (V$_{\text{expt}}$) and theoretical volumes (V$_{\text{theo}}$) respectively. }
	\label{crystal_TI}
\end{figure*}
Cu$_2$ZnGeTe$_4$ crystallizes in a tetragonal lattice with a non-centrosymmetric space group I$\bar{4}$2m ($\#$121 hosting a D$_{2d}$ point group), as shown in Fig. \ref{crystal_TI}(a) and confirmed by our X-ray diffraction data (see Sec. III(B) for further experimental details). It has two formula units, leading to 16 atoms in the unit cell. This structure can also be visualized as the supercell of the zinc-blende structure along the z-direction where three different types of cations (Cu, Zn, Ge) are substituted at the zinc site and sulphur is replaced by tellurium, and is often known as the low-symmetry non-cubic structure. Every atom has a tetrahedral surrounding, and the tetrahedron is made of anionic tellurium and distinct cationic metal cations (see Fig. \ref{crystal_TI}(a)). This difference in electronegativity between the cations and the different interatomic distances leads to a naturally occurring superlattice structure with localised lattice distortions. Notably, the bond lengths between these three different cations and the Te anions are 2.695 \AA, 2.691 \AA, 2.576 \AA \space for Zn-Te, Ge-Te and Cu-Te, respectively. Thus, it forms a locally distorted kind of anionic framework.

\textcolor{black}{Table \ref{lattice} compares the experimentally measured and theoretically optimized lattice parameters of Cu$_2$ZnGeTe$_4$. The PBE-optimized structure has a larger unit-cell volume than the experimentally determined structure and may therefore be regarded as a strain-expanded configuration relative to the experimentally realised phase. Such a modest lattice expansion ($\sim$ 6\%) produces a pronounced modification of the electronic structure. In the following discussion, the structures based on the measured and expanded lattice parameters are referred to as the experimental-volume semiconducting phase and the strain-expanded phase, respectively.}
%\textcolor{red}{Trying to seperate the experimental-theory part from topological part (only theory) so gave a nomenclature to both the phases.}
The computed crystal orbital Hamiltonian population (COHP) for the interaction between the Te -(Cu/Zn/Ge) elements reveals the distinct bonding nature at these two volumes. Figure 2(c,d) demonstrates how distinct cations and anions in Cu$_2$ZnGeTe$_4$ show different energy-dependent bonding (negative) and antibonding (positive) features at experimental and theoretical volumes, respectively. The COHP curve for the theoretical volume shows antibonding states at the E$_F$, while there are no antibonding states at E$_F$ for experimental volume, indicating a metallic ground state at the theoretical volume (see insets in Figs. \ref{crystal_TI}(c,d)) The locally distorted structure of this compound might be one of the reasons as specified above. In the following sections, we will report a detailed electronic/thermal transport properties of Cu$_2$ZnGeTe$_4$ at both the experimental and theoretical volumes and discuss the contrasting features. Later, we will compare the trend of our theoretical results with those measured in Sec. III(B).

\subsubsection{Electronic structure}
\begin{figure}[t]
	\centering
	\includegraphics[scale=0.5]{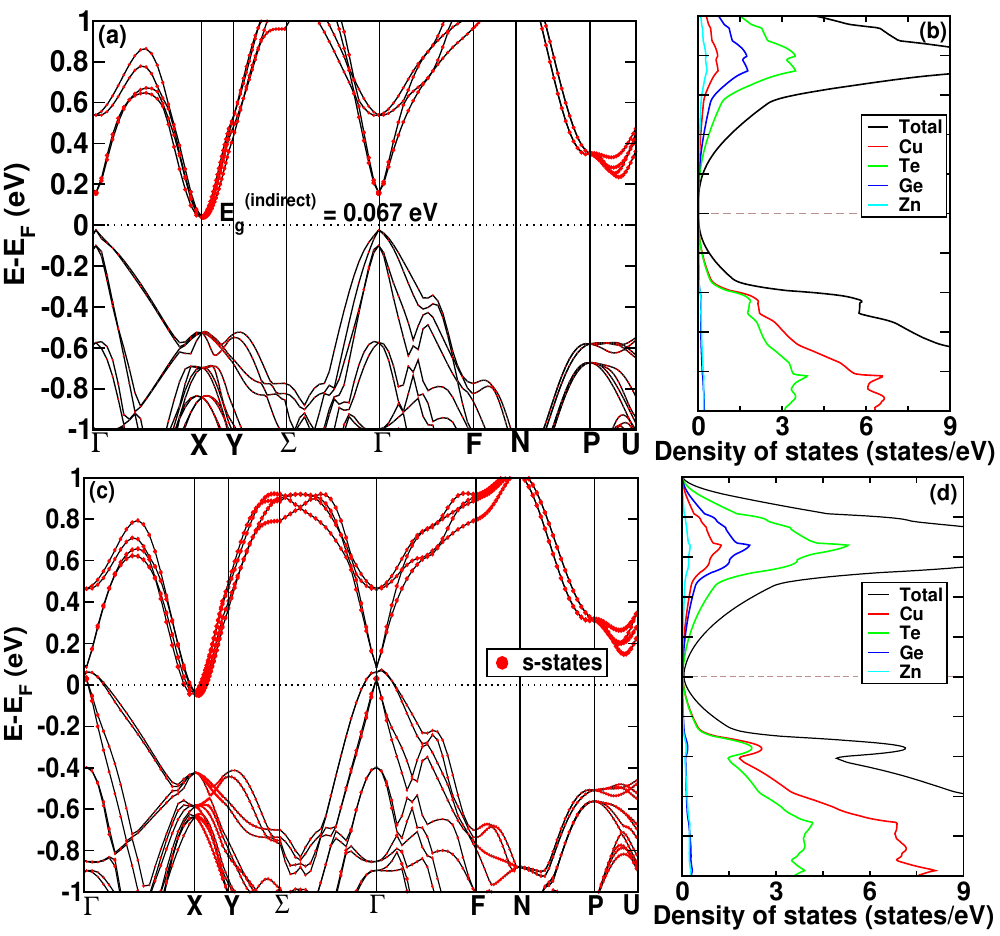}
	\caption{For Cu$_2$ZnGeTe$_4$, electronic band structure and atom-projected density of states at (a,b) experimental volume (c,d) theoretically optimized volume. The former is a narrow band gap semiconductor while the latter is a Weyl semi-metal. Red circles indicate s-states. Fermi level (E$_F$) is set at 0 eV.}
	\label{EB_ap}
\end{figure}

 Figure \ref{EB_ap}(a,b) shows the band structure and density of states of Cu$_2$ZnGeTe$_4$ at experimental lattice parameters. It shows an indirect band gap of 0.067 eV. The most dominant contribution to the conduction bands arises from Te p-orbitals while that to valence bands arise from Cu d- and Te p-orbitals. Conduction band minima (CBM) occur at X point, while valence band maxima (VBM) occur at $\Gamma$ point making it an indirect band gap semiconductor. There is yet another minima in the conduction band at $\Gamma$ point slightly above the one at X-point. The low band gap and the presence of multiple bands near the Fermi level are the preliminary signatures of promising transport properties. 
\begin{figure}[!b]
	\centering
	\includegraphics[scale=0.26]{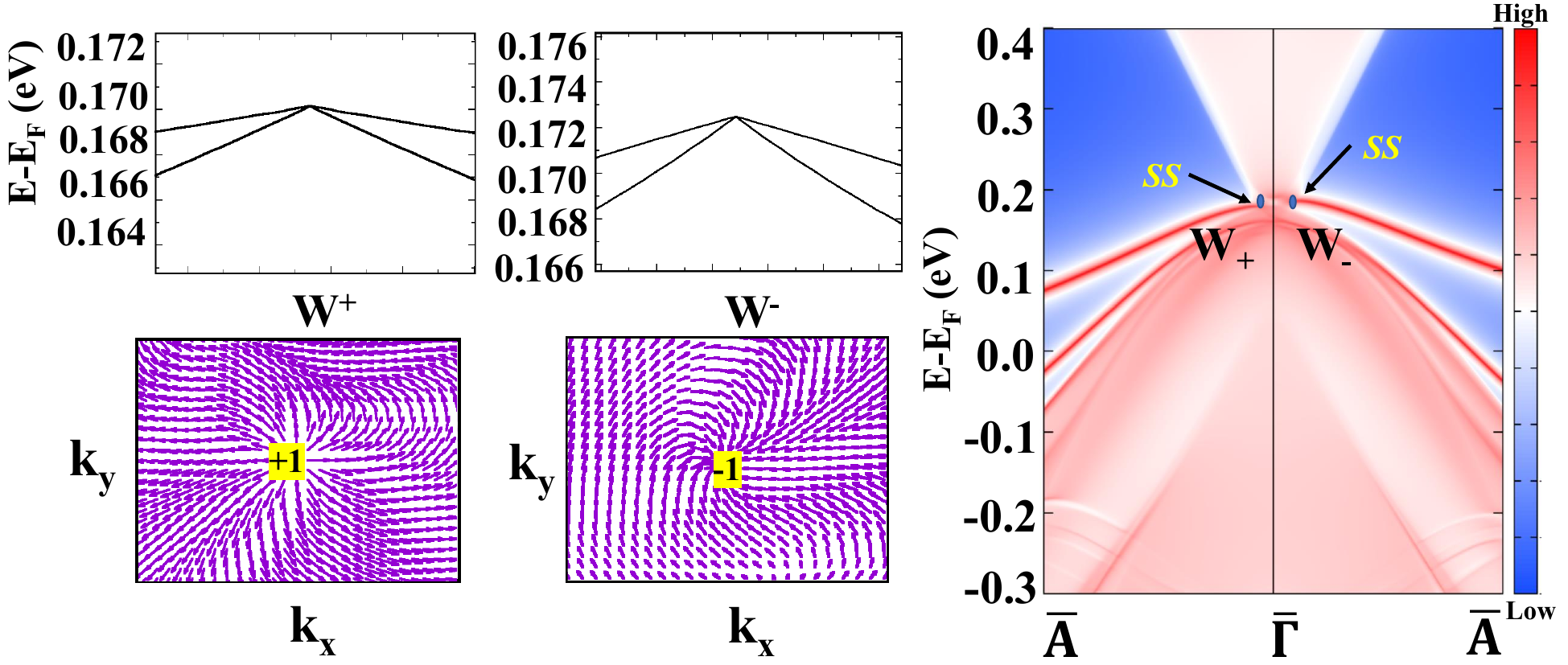}
	\caption{For Cu$_2$ZnGeTe$_4$ (in semi-metallic state), (a,b) zoomed-in view of bulk band structure around Weyl points W$^+$ and W$^-$ with chirality +1 and -1 respectively (c,d) Berry curvature around W$^+$ and W$^-$ (e) surface band structure projected along (001) direction.}
	\label{WP}
\end{figure}

In contrast, Cu$_2$ZnGeTe$_4$ shows a semi-metallic feature at the theoretically optimised volume, as shown in Fig. \ref{EB_ap}(c). Interestingly, it also exhibits a band inversion around $\Gamma$ point originating from the flipping of 's' character, indicating the possibility of topological non-trivial nature. \textcolor{black}{Relative to the experimental-volume structure, lattice expansion closes the semiconducting gap and produces an inversion of the relevant bands around the $\Gamma$ point and thus driving a transition from a topologically trivial semiconductor to a topologically nontrivial Weyl semimetal.} Broken inversion symmetry confirms the possibility of Weyl nodes, provided it is a topologically non-trivial semimetal. A search in the entire Brillouin zone (BZ) revealed the presence of a pair of Weyl points near the $\Gamma$ point labelled as W$^{+}$ and W$^{-}$ (with chirality +1 and -1), whose coordinates are given in Table \ref{corr}. The bulk band dispersion around these two Weyl points are shown in Fig. \ref{WP}(a,b). These Weyl points have opposite chirality and obey the Nielsen-Ninomiya theorem.\cite{NN_theorem} The location of these Weyl points is slightly above the Fermi level as shown in the zoomed-in Figure \ref{WP}(a,b). 
\begin{table}[t]
	\centering{}
	\begin{tabular}{|c|c|c|c|c|}
		\hline 
		Points & k$_x$ & k$_y$  & k$_z$ & Chirality \tabularnewline
		\hline 
		\hline 
		W$^+$ 	&\ \ \ 0.00155\ \ \ &\ \ \ 0.00693\ \ \ &\ \ \ 0.02230\ \ \  &+1 \tabularnewline \hline 
		W$^-$ &\ \ \ 0.00794\ \ \ &\ \ \ -0.00320\ \ \ &\ \ \ -0.02663\ \ \ &-1\tabularnewline \hline 
	
		\hline 		
	\end{tabular}
	\caption{For Cu$_2$ZnGeTe$_4$ (in semi-metallic state), coordinates of Weyl points in the reciprocal space with opposite chirality.}
	\label{corr}
\end{table}
The corresponding Berry curvature around these points showing the source and sink nature are presented in Fig. \ref{WP}(c,d). Surface states are very important feature for a topological material. Since, this compound shows Weyl nodes in the bulk band structure, it is expected to host a robust surface state originating from this bulk Weyl point. 
The surface band structure projected on the (001) surface is shown in Figure \ref{WP}(e) which clearly shows two Weyl points at/around $\Gamma$ point. 

 \begin{figure*}[!t]
\centering
	\includegraphics[scale=0.16]{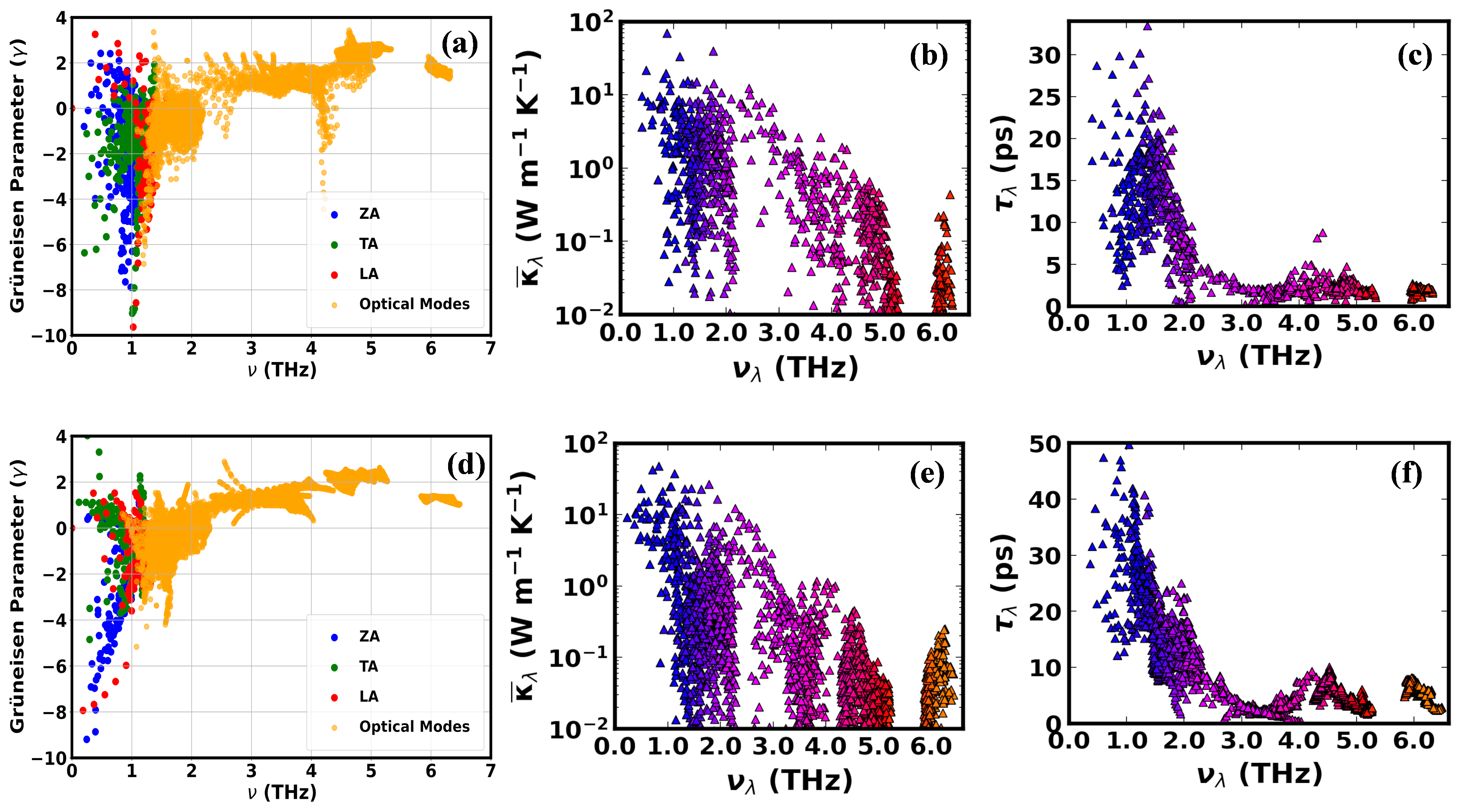}
\caption{(a,b,c) The mode Gruneisen parameter ($\gamma$), frequency dependent lattice thermal conductivity ($\kappa_{\lambda}$) and phonon relaxation time ($\tau_{\lambda}$) for semi-conducting phase of Cu$_2$ZnGeTe$_4$ (d,e,f) Same as (a,b,c) but for semi-metallic state.}
\label{fig1}
\end{figure*}
\begin{figure}[!b]
\centering
		\includegraphics[scale=0.3]{kappa.png} 
\caption{The total, in-plane and out-of-plane lattice thermal conductivity ($\kappa_L$) vs. T for semi-conducting and semi-metallic state of Cu$_2$ZnGeTe$_4$. }
\label{fig2}
\end{figure}

\subsubsection{Phonon properties}

Phonons play a crucial role in dictating various novel properties of a material. Cu$_{2}$ZnGeTe$_{4}$ contains 2 formula units including 16 atoms in the unit cell leading to a total of 48 bands in the phonon spectrum. This is shown in Fig. SI(a,b) of SM\cite{suppl} for both the semi-metallic and semi-conducting phases. Among them, three bands originate from acoustic phonon branches while the remaining 45 bands represent the optical phonon modes. Interestingly this compound facilitates various band crossings in both acoustic and optical phonon bands. A close observation of the acoustic region reveals highly entangled acoustic and low-frequency optical modes, which eventually helps to reduce lattice thermal conductivity and are hence beneficial for thermoelectric (TE) applications. The atom projected phonon density of states shown in Fig. SI(c,d)\cite{suppl} helps to identify the contribution of different atoms in different frequency ranges of the phonon spectra. The heavy element `Te' contributes in the low frequency acoustic region while the other three elements shows up at a relatively higher frequency range. 
%For a closer inspection, the phonon dispersion for semi-conducting state in the low frequency range is shown in Fig. SII\cite{suppl}.\textcolor{magenta}{[For wihch phase, semiconducting or semimetallic? why are you not showing the same figure for the other phase ?]} The eigenmode displacements of few of the low-frequency modes are shown in Fig. SIII\cite{suppl} . The predominant vibrations arises from Te-atoms, followed by Cu. The mixing of the acoustic and optical phonon modes can be seen in 3$^{rd}$, 4$^{th}$, 5$^{th}$ and 6$^{th}$ bands at X, N and P high symmetry points. 

The change in phonon mode frequencies with slight change in volume is explained by the frequency dependence of mode Grüneisen parameter ($\gamma$), as shown in Fig. \ref{fig1}(a,d) for semi-conducting and semi-metallic phases respectively. $\gamma$ is a measure of the anharmonicity of the lattice; the higher the $\gamma$-value, the more anharmonic the lattice is; which causes more scattering and hence lowers the $\kappa_L$ values. At low frequencies, phonons, particularly the acoustic modes, show negative values of mode gruneisen parameter which could be due to the special structure framework where the anions are at the middle of the tetrahedral void and the corners are shared by cations. At low frequencies, transverse displacement of the anion (Te) atom causes other cationic atoms in the void to shrink towards each other.\cite{neg_grun} Figures \ref{fig1}(b,c) and (e,f) show the frequency-dependent cumulative lattice thermal conductivity and phonon relaxation time for the semiconducting and semimetallic phases, respectively. From Fig. \ref{fig1}(b,d), it is evident that lattice thermal conductivity is predominantly contributed by the acoustic phonon modes, i.e., up to around 2 THz in both phases. Figure \ref{fig1}(c,f) further indicates that with increasing frequency, the relaxation time decreases due to enhanced phonon–phonon scattering arising from the mixing of different phonon branches at higher frequencies. A comparison between the semiconducting and semimetallic phases reveals that the semimetal exhibits a lower scattering rate, which in turn leads to a higher lattice thermal conductivity. Figure \ref{fig2} compares the lattice thermal conductivity ($\kappa_L$) for semi-conducting and semi-metallic state of Cu$_{2}$ZnGeTe$_{4}$. Semiconducting phase shows lower values of $\kappa_L$ because of a more entangled acoustic and optical phonon branches in the low frequency range, as well as a more compact (locally) distorted anionic framework. 

%\textcolor{magenta}{\Large Until here !!!!}
\begin{figure*}[t]
	\centering
	\includegraphics[scale=0.62]{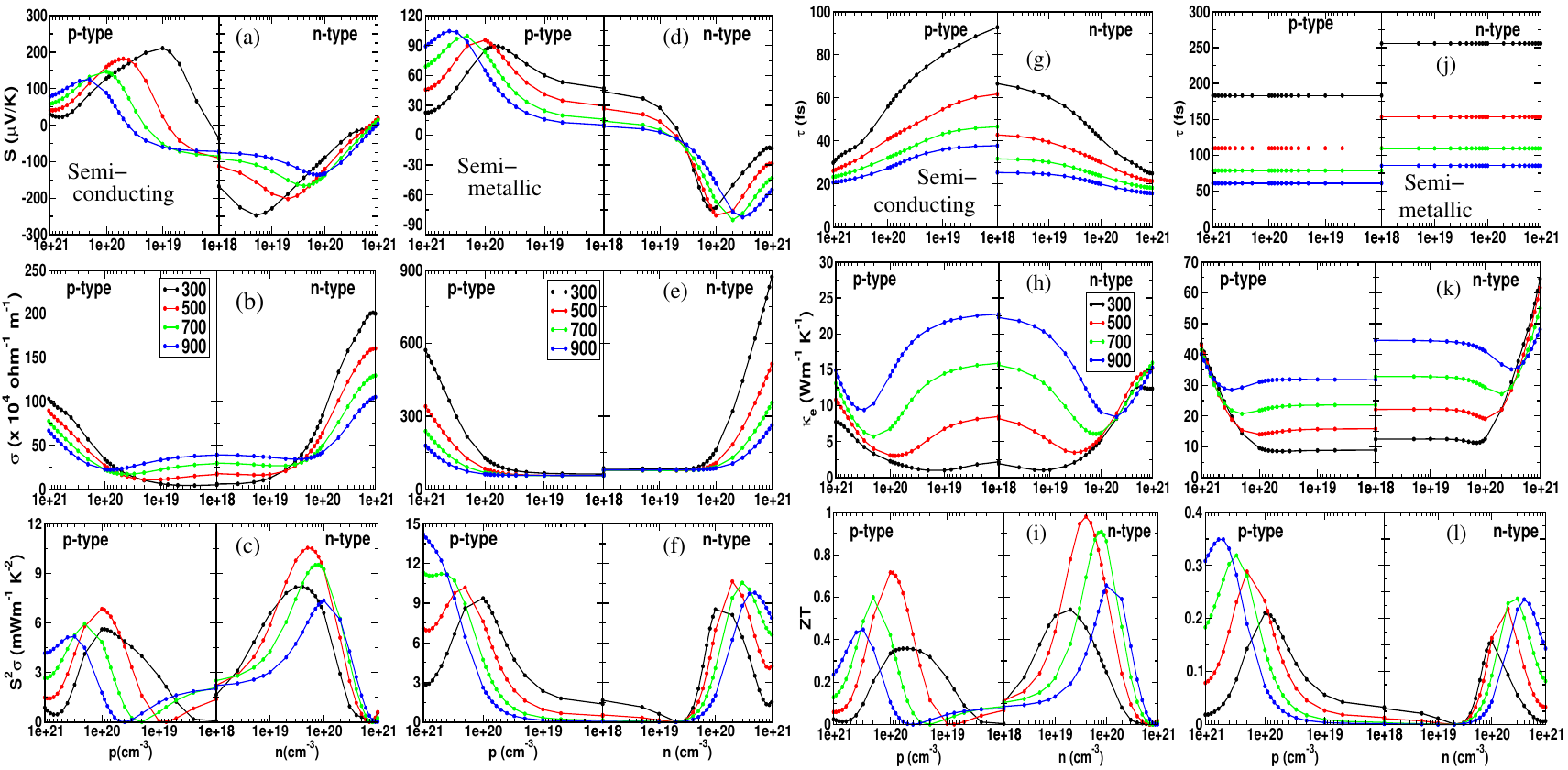}
	\caption{For Cu$_2$ZnGeTe$_4$, carrier concentration dependence of (a,d) Seebeck coefficient (S), (b,e) $\sigma$ (c,f) S$^2\sigma$, (g,i) $\kappa_e$ and (h,j) ZT at different T for  (left) p-type and (right) n-type conduction for semiconducting (left) and semi-metallic (right) states respectively. }
    \label{ZT}
\end{figure*}

\subsubsection{Thermoelectric properties}
Band topology plays a crucial role in governing the transport properties, and in turn, the TE performance of a material. Band convergence\cite{BE1} is a clever strategy to improve TE performance which is quite evident in Cu$_2$ZnGeTe$_4$. There are regular parabolic bands in both the conduction and valence side of the band structure along with the light bands (see Figure \ref{EB_ap}). This maximizes both Seebeck coefficient and electrical conductivity. Along with that, low values of thermal conductivity further helps to enhance the TE figure of merit. In the following subsections, we will describe the thermoelectric (TE) properties of Cu$_2$ZnGeTe$_4$ in both semi-conducting (experimental volume) and semi-metallic (theoretical volume) states. 

\textbf{Semi-conducting state:} The transport properties are calculated by considering different scattering effects namely, acoustic deformation potential (ADP), ionised impurity (IMP), polar optical phonon (POP) and piezoelectric (PIE) scattering within the ab-initio framework.\cite{AMSET} The individual relaxation times due to ADP, POP, IMP scattering mechanisms and the total relaxation time ($\tau$)  for semiconducting Cu$_2$ZnGeTe$_4$ at different values of carrier concentrations and temperatures for p-type and n-type conduction are shown in Fig. SII of SM.\cite{suppl} The PIE scattering rates are negligibly small in the temperature range 300-900 K, and hence are not shown. The value of $\tau$ at room temperature is around $\sim$ 80 fs for holes and $\sim$ 60 fs for electrons at a carrier concentration of 1$\times$ 10$^{19}$ cm$^{-3}$, which decreases to 36 fs and 25 fs at 900 K for holes and electrons respectively. Figures \ref{ZT}(a,b,c) shows the variation of Seebeck coefficient (S), electrical conductivity ($\sigma$), and power factor (S$^2$$\sigma$) with respect to carrier concentration at different temperatures for p-type (left) and n-type (right) conduction for semiconducting Cu$_2$ZnGeTe$_4$. The higher $\sigma$ values ($\sim$ 2$\times$10$^6$ S/m for n-type conduction) are due to the linearly dispersive nature of conduction bands. These values are quite high as compared to other copper-based quaternary chalcogenides (Cu$_2$B$^{\text{II}}$C$^{\text{IV}}$X$_4$) which show conductivity in the range 10$^3$ - 10$^4$ S/m. This, together with the large S values, results in a high power factor of 10.57 mWm$^{-1}$K$^{-2}$ at 500 K. This competes with the reported power factor values of well-known promising thermoelectric materials. The power-factor values of p-type are relatively lower with the maximum value around 6.9 mWm$^{-1}$K$^{-2}$ at 500 K. 

Figure \ref{ZT}(g,h,i) shows the variation of total relaxation times ($\tau$), electronic thermal conductivity ($\kappa_e$) and thermoelectric figure of merit (ZT) as a function of carrier concentration at different temperatures for p-type (left) and n-type (right) conduction for semiconducting Cu$_2$ZnGeTe$_4$. ADP is the dominant scattering mechanism in the carrier concentration range of 10$^{18}$-10$^{20}$ cm$^{-3}$ (see Fig. SII of SM). IMP becomes dominant at higher concentrations. Due to the low value of bandgap (E$_g$), this material shows bipolar transport even at 300 K. The higher $\kappa_e$ values and lower S values at lower concentrations are due to bipolar conduction. \cite{bipolar} This happens when minority carriers also start contributing to the transport.
The maximum value of ZT in the semiconducting phase is $\sim$ 0.7 and $\sim$ 1.0 for p-type and n-type conduction respectively at 500 K. \textcolor{black}{At a fixed value of carrier concentration of 1 $\times$ 10$^{20}$ c at which ZT is maximum for p-type, the values of S, $\sigma$ and S$^2\sigma$ are 159.6 $\mu$V/K, 2.69$\times$10$^5$ $\Omega^{-1}$m$^{-1}$ and 6.86 mWm$^{-1}$K$^{-2}$ respectively. The same for n-type conduction are 179 $\mu$V/K, 3.21$\times$10$^5$ $\Omega^{-1}$m$^{-1}$ and 10.35 mWm$^{-1}$K$^{-2}$ at a carrier concentration of 4$\times$10$^{19}$ cm$^{-3}$.}

 \textbf{Semi-metallic state}
 Topological semi-metals such as Weyl/Dirac materials can offer a perfect combination of light and heavy bands which can yield a large $\sigma$ as well as a promising Seebeck coefficient (S), enabling optimum power-factor (S$^2\sigma$). Since ADP is the dominant scattering mechanism in a wide concentration range of 10$^{18}$-10$^{20}$ cm$^{-3}$ in the semi-conducting phase, we estimated the carrier relaxation time for the semi-metallic phase by considering the ADP scattering mechanism only. The total relaxation times as a function of carrier concentration at different temperatures for p-type and n-type conduction are shown in Fig. \ref{ZT}(j). The values of $\tau$ at room temperature are around $\sim$ 180 fs for holes and 256 fs for electrons respectively. This decreases to 60 fs and 85 fs at 900 K for both holes and electrons respectively. Figures \ref{ZT}(d,e, and f) show the variation of S, $\sigma$, and S$^2\sigma$ with respect to carrier concentration at different temperatures for p-type and n-type conduction of semi-metallic Cu$_2$ZnGeTe$_4$. The asymmetry in the density of states (Fig. \ref{EB_ap}(d)) around the Fermi level results in a Seebeck coefficient value of $\sim$ 50 $\mu$V/K at a low concentration of 1 $\times$ 10$^{18}$ (absence of doping) at room temperature, which peaks to $\sim$ 90 $\mu$V/K at a carrier concentration of $\sim$ 7 $\times$ 10$^{19}$ cm$^{-3}$ for p-type (see Fig. \ref{ZT}(d)). The $\sigma$ values for electrons in the present case are approximately 4 $\times$ higher than the semi-conducting phase due to higher values of relaxation times.
 The high values of power-factor ($\sim$ 9 mWm$^{-1}$K$^{-2}$) are obtained for both p-type and n-type at room temperature which peaks to 10.6 mWm$^{-1}$K$^{-2}$ at 500 K for n-type and 14.2 mWm$^{-1}$K$^{-2}$ at 900 K for p-type due to higher S values at 900 K for p-type.

Figures \ref{ZT}(k,l) show the variation of $\kappa_e$, and ZT as a function of carrier concentration at different temperatures for p-type and n-type conduction of semi-metallic Cu$_2$ZnGeTe$_4$. At room temperature, and at a fixed carrier concentration of 1$\times$10$^{20}$ cm$^{-3}$, the values of S, $\sigma$ and S$^2\sigma$ are 85.5 $\mu$V/K, 12.8 $\times$ 10$^5$ $\Omega^{-1}$m$^{-1}$ and 9.4 mWm$^{-1}$K$^{-2}$ respectively leading to a ZT value of 0.21 for p-type. When compared with other semi-metals reported so far, this turn out to be a reasonably large ZT value at room temperature. At higher temperatures, the ZT value further increases to $\sim$ 0.35 for p-type conduction at a carrier concentration of 5 $\times$10$^{20}$cm$^{-3}$ as shown in Fig. \ref{ZT}(l). For n-type conduction, the maximum ZT value achieved at room temperature and at carrier concentration of 1 $\times$ 10$^{20}$ cm$^{-3}$ is $\sim$ 0.16 which further increases to $\sim$ 0.24 at 800 K at a carrier concentration of 4$\times$10$^{20}$ cm$^{-3}$. 
\textcolor{black}{Although the strain-expanded Weyl phase exhibits high electrical conductivity and a large power factor, its maximum ZT remains lower than that of the semiconducting phase. The enhancement of $\sigma$ simultaneously increases the electronic thermal conductivity through the Wiedemann–Franz relation, therefore, opposing most of the gain from the high power-factor. In addition, the expanded phase possesses a higher lattice thermal conductivity because of its longer phonon relaxation times and reduced acoustic–optical mode mixing. The Weyl transition is therefore favourable for electronic transport but not for the total thermoelectric efficiency. Its significance lies in maintaining a competitive power factor and a moderate $ZT$ following the emergence of nontrivial band topology, rather than in an improvement of ZT over the parent semiconductor. In Sec. IV, we investigate Sn substitution as an alternate chemical route for producing the required lattice expansion and accessing the topologically nontrivial phase.}
%\textcolor{red}{Trying to emphasize on achievement in transport for topological phase and discussing the bottleneck.}

\subsection{Experimental Results}
The X-ray diffraction pattern [Fig.~SIV(a)] confirms the formation of single-phase Cu$_2$ZnGeTe$_4$ crystallising in the tetragonal structure with space group I$\bar{4}$2m. The relatively broad diffraction peaks are indicative of nanostructured grains, originating from the ball-milling process employed during sample preparation (see Supporting Information for synthesis details). Such peak broadening is commonly associated with nanostructuring and microstrain induced by severe mechanical deformation during milling. The thermal stability of the compound is verified by thermogravimetric analysis [Fig.~SIV(b)], which reveals no significant weight loss up to 640~K. The experimentally estimated band gap of approximately 0.11~eV [Fig.~SIV(c)] is in good agreement with the calculated value of 0.067~eV. The small discrepancy can be attributed to the well-known tendency of semilocal GGA functionals to underestimate band gaps, as well as possible effects arising from the nanostructured nature of the synthesised sample.

\begin{figure}[t]
	\centering
	\includegraphics[scale=0.12]{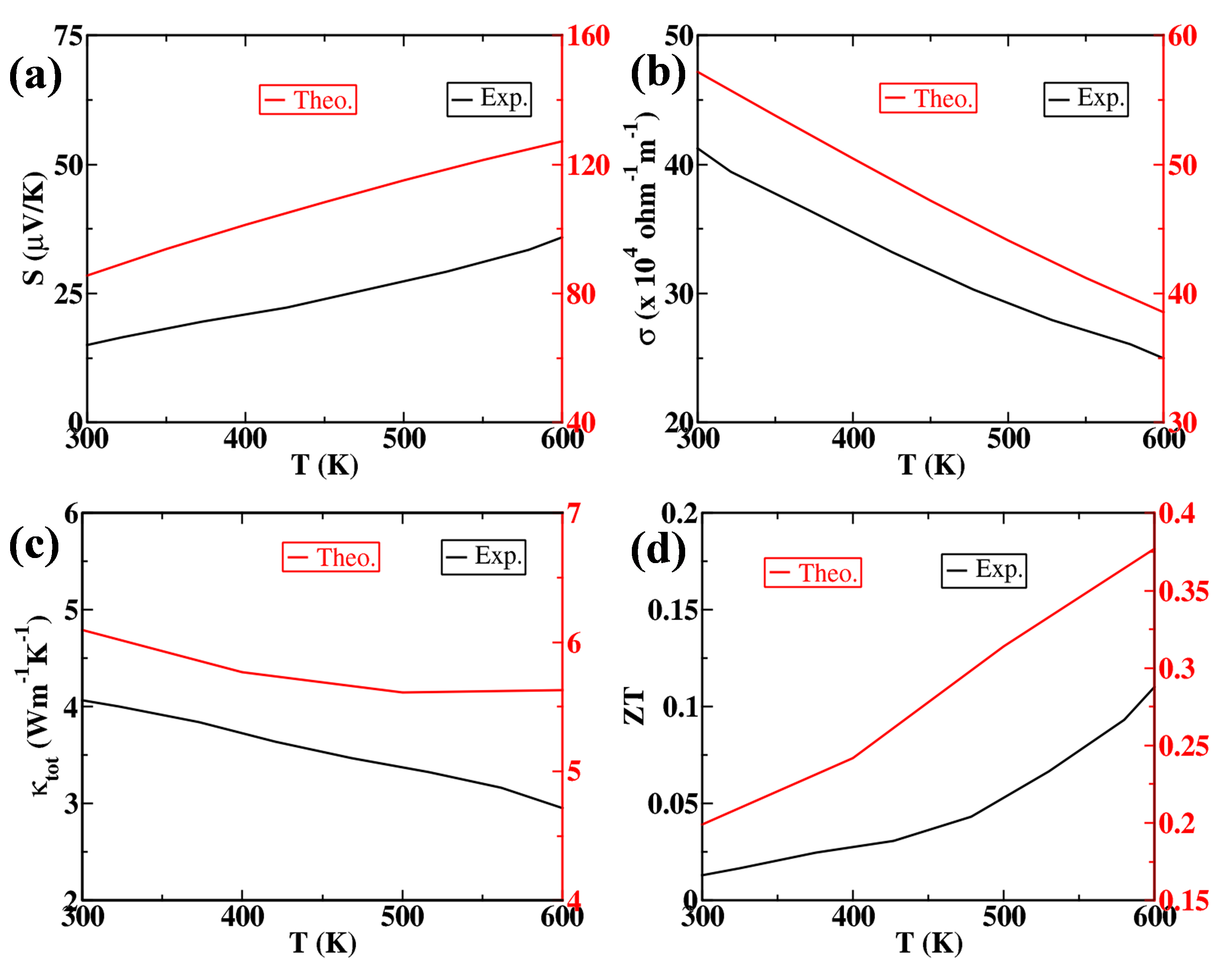}
	\caption{Comparison of theoretical and experimental results for temperature (T) dependence of (a) Seebeck coefficient (S) (b) electrical conductivity ($\sigma$), (c) total thermal conductivity ($\kappa_{tot}=\kappa_e$+$\kappa_L$) and (d) TE figure of merit (ZT) of Cu$_2$ZnGeTe$_4$ at an optimised carrier concentration of $2 \times 10^{20}~\mathrm{cm}^{-3}$. These theoretical results are obtained using the experimental lattice parameters. }
	\label{compare_ap}
\end{figure}

A direct comparison between the measured and calculated transport properties is presented in Fig.~\ref{compare_ap}(a--d). The simulated results correspond to a hole concentration of $2 \times 10^{20}~\mathrm{cm}^{-3}$, a carrier-density regime in which the system exhibits degenerate semiconducting behaviour consistent with experimental observations. As shown in Fig.~\ref{compare_ap}(b), the electrical conductivity ($\sigma$) decreases monotonically with increasing temperature in both theory and experiment, a hallmark of degenerate charge transport. Although the calculated conductivity is quantitatively larger, the experimentally observed temperature dependence is reproduced remarkably well. The higher simulated values are primarily attributed to the absence of grain-boundary and defect scattering in the idealised crystal model, whereas such scattering mechanisms significantly reduce carrier mobility in the ball-milled polycrystalline samples.

Figure~\ref{compare_ap}(a) shows the temperature (T) dependence of Seebeck coefficient [$S$], which remains positive throughout the investigated  T-range, confirming the dominant $p$-type character of charge transport. While the measured values are somewhat lower than the theoretical predictions, both exhibit a nearly linear increase with temperature. \textcolor{black}{The sizeable thermopower can be understood from the underlying electronic structure, where relatively flat parabolic bands contribute a large density of states effective mass, thereby enhancing $S$, whereas the coexisting dispersive bands support comparatively high carrier mobility.} This interpretation is consistent with Hall-effect measurements, which reveal a mobility of approximately $448~\mathrm{cm}^{2}\mathrm{V}^{-1}\mathrm{s}^{-1}$.\cite{exp1,exp2,exp3,exp4}

Figure~\ref{compare_ap}(c) shows the total thermal conductivity ($\kappa_{\mathrm{tot}}$) vs. temperature. $\kappa_{\mathrm{tot}}$ is found to decrease gradually with increasing temperature from 300 to 600~K. The simulated values exceed the measured ones, which is consistent with the higher predicted electrical conductivity and the corresponding enhancement of the electronic thermal conductivity ($\kappa_e$). Notably, the experimentally extracted lattice thermal conductivity ($\kappa_L$) reaches an ultralow value of $0.37~\mathrm{W\,m^{-1}K^{-1}}$ at 623~K. Such strong suppression of heat transport is in fair agreement with our first-principles phonon calculations, which reveal pronounced anharmonicity and significant hybridisation between acoustic and low-frequency optical phonon modes. These results indicate that the exceptionally low $\kappa_L$ is an intrinsic consequence of the crystal chemistry of Cu$_2$ZnGeTe$_4$.\cite{exp5}

\textcolor{black}{The combined influence of the electronic and thermal transport properties is reflected in the dimensionless thermoelectric figure of merit $ZT$, shown in Fig.~\ref{compare_ap}(d). Experimentally, $ZT$ reaches approximately 0.14 at 623 K, whereas the calculations performed at the experimental lattice parameters predict a value of approximately 0.4 over the same temperature range. Despite this quantitative difference, the experimental data closely follow the theoretically predicted temperature dependence. The lower measured $ZT$ primarily originates from the reduced electrical conductivity and Seebeck coefficient caused by microstructural scattering effects such as grain boundaries, defects and microstrain  introduced during synthesis and ball milling provide additional carrier-scattering channels that are not explicitly included in the calculations. The experimentally measured ultralow lattice thermal conductivity further reflects both the intrinsic anharmonicity of Cu$_2$ZnGeTe$_4$	and additional phonon scattering from its nanostructured microstructure. The present measurements therefore establish the ambient Cu$_2$ZnGeTe$_4$ phase as an experimentally accessible, narrow-gap thermoelectric semiconductor. Having established the experimental relevance of the parent compound, we next investigate Sn alloying as a possible chemical route toward the expanded topologically nontrivial phase.}
%\textcolor{red}{Trying to make sure that this section does not indicate the overlap between semiconducting and topological part. I mean experiment is only done for one phase.}

\begin{figure}[t]
	\centering
	\includegraphics[scale=0.66]{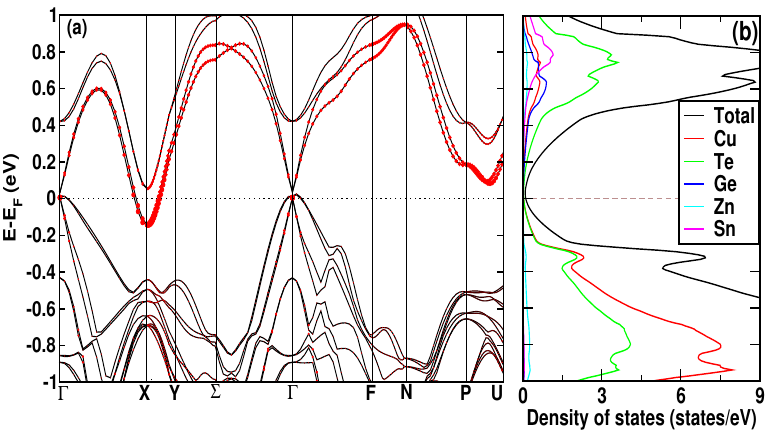}
	\caption{(a) Electronic band structure and (b) total/atom-projected density of states of Cu$_2$ZnGe$_{0.5}$Sn$_{0.5}$Te$_4$.}
	\label{doped}
\end{figure}

\section{Alloy engineering}
\textcolor{black}{We next considered isovalent Sn substitution at the Ge site as a chemical route for producing the lattice expansion required to access the topologically nontrivial phase. The substitutional site was chosen carefully so as not to affect the band topology much at/near the E$_F$. Ge is an ideal site whose contribution at/near the E$_F$ is small (see the density of states plots in Fig. \ref{EB_ap}(b,d)). Replacing 50\% of Ge by the larger Sn atom produces Cu$_2$ZnGe$_{0.5}$Sn$_{0.5}$Te$_4$, whose theoretically optimized unit-cell volume is 461.9 \AA$^3$. The calculated band structure, shown in Fig. \ref{doped}, retains the combination of relatively flat and dispersive bands near the Fermi level while displaying the band inversion associated with the expanded structure. The fourfold-degenerate conduction-band minimum at the X point of the parent compound (Fig. \ref{EB_ap}(a,c)) splits into two doubly degenerate bands after alloying. These results show that chemical expansion through Sn substitution can reproduce the essential electronic features of the strain-expanded phase and provides a possible route for accessing the predicted topologically nontrivial state without externally applied strain.}
%\textcolor{red}{Just imporved the terminology in this section}

\section{Conclusion}

We have investigated the structural, electronic, topological, phononic, and thermoelectric properties of Cu$_2$ZnGeTe$_4$ and established a direct connection between lattice tuning, electronic topology, and thermoelectric transport. At the experimentally measured lattice parameters, Cu$_2$ZnGeTe$_4$ is a narrow-gap semiconductor with a calculated band gap of approximately 0.067~eV and a maximum calculated $ZT$ of approximately 1 for $n$-type transport. Our experimental synthesis confirms the $p$-type semiconducting character of the parent compound and reveal an ultralow lattice thermal conductivity, establishing its potential as a thermoelectric material. The lower experimental $ZT$ compared with the ideal-crystal prediction is primarily associated with reduced electronic transport in the polycrystalline samples.

Lattice expansion drives a band inversion and transforms the electronic structure into a Weyl-semimetallic phase hosting Weyl nodes of opposite chirality, associated Berry-curvature source and sink features, and topological surface states. The Weyl phase exhibits enhanced electrical conductivity and a high power factor arising from its dispersive electronic states and longer carrier relaxation times. However, the accompanying increases in both electronic and lattice thermal conductivities offset these advantages, limiting the maximum $ZT$ to approximately 0.36. Thus, while Weyl topology can coexist with favourable electronic transport and a competitive power factor, it does not by itself guarantee enhanced thermoelectric efficiency.

Importantly, Cu$_2$ZnGeTe$_4$ demonstrates that nontrivial electronic topology can be accessed without completely sacrificing thermoelectric performance. Isovalent Sn substitution at the Ge site provides a possible chemical-expansion route toward the predicted topologically nontrivial phase. Overall, Cu$_2$ZnGeTe$_4$ offers a tunable platform for exploring the interplay between topological electronic structure and thermoelectric transport, while highlighting the need to simultaneously engineer electronic and phononic properties to realise high-performance topological thermoelectric materials.

\section{Acknowledgements} B.S. acknowledges the computing facilities (spacetime) of IIT Bombay to carry out the calculations. H.S. acknowledges the financial support from the Indian Institute of Technology, Bombay, in the form of a teaching assistantship. H.S. also acknowledge the support of HPC facility PARAM Rudra under the National Supercomputing Mission. A.A. acknowledges DST-SERB (Grant No. CRG/2019/02050) for funding to support this research.

\bibliographystyle{unsrtnat}
\bibliography{main.bib}

\end{document}